\documentclass[prl,aps,superscriptaddress]{revtex4}
\usepackage{bm}
\usepackage{amsfonts}
\usepackage[dvips]{graphicx}
\usepackage{mathrsfs}
\usepackage[intlimits]{amsmath}
\usepackage[colorlinks, citecolor=red]{hyperref}
\usepackage{color}

\begin{document}

\title{Experimental entanglement of $25$ individually accessible atomic
quantum interfaces}
\author{Y.-F. Pu$^{1}$, Y.-K. Wu$^{1,2}$, N. Jiang$^{1}$, W. Chang$^{1}$, C.
Li$^{1}$, S. Zhang$^{1}$, L.-M. Duan}
\affiliation{Center for Quantum Information, IIIS, Tsinghua University, Beijing 100084,
PR China}
\affiliation{Department of Physics, University of Michigan, Ann Arbor, Michigan 48109, USA}
\begin{abstract}
{A quantum interface links the stationary qubits in a quantum memory with flying
photonic qubits in optical transmission channels and constitutes a critical
element for the future quantum internet. Entanglement of quantum interfaces is an
important step for the realization of quantum networks. Through heralded detection of
photon interference, here we generate multipartite entanglement between $25$
(or $9$) individually addressable quantum interfaces in a multiplexed atomic
quantum memory array and confirm genuine $22$ (or $9$) partite entanglement,
respectively. This experimental entanglement of a
record-high number of individually addressable quantum
interfaces makes an important step towards the realization of quantum
networks, long-distance quantum communication, and multipartite quantum
information processing.}
\end{abstract}
\maketitle

\section{Introduction}

Stationary qubits carried by the ground states of cold atoms are an ideal memory
for storage of quantum information, while flying photonic pulses are the
best choice for the transmission of quantum information through the optical
communication channels. A quantum interface can convert the stationary qubits
into the flying photonic pulses and vice versa, and therefore generates an
efficient link between the quantum memory and the optical communication channels \cite%
{hammerer2010quantum}. A good quantum memory is provided by the hyperfine states of single atoms (ions)
or the collective states of an atomic ensemble. Compared with single atoms or ions, the collective state
of an atomic ensemble cannot be easily controlled for performing qubit rotations
and qubit-qubit gate operations, and therefore it is not a convenient qubit for the realization of quantum computation.
However, due to the collective enhancement effect, the collective state of an optically dense atomic ensemble has a unique advantage of strong coupling to the directional emission even in the free space, which generates an efficient quantum link
between the atomic memory and the forward-propagating photonic pulses and hence provides an ideal candidate for the realization of the
quantum interface  \cite%
{duan2001long,hammerer2010quantum,kimble2008quantum}. For implementation of quantum networks,
long-distance quantum communication, and the future quantum internet,
a promising way of scaling up is based on generating entanglement between these efficient
quantum interfaces \cite
{briegel1998quantum,duan2001long,kimble2008quantum,hammerer2010quantum,sangouard2011quantum,collins2007multiplexed}
. Remarkable experimental advances have been reported towards this goal \cite%
{chou2005measurement,chaneliere2005storage,eisaman2005electromagnetically,julsgaard2004experimental,simon2007single,de2008solid,
lan2009multiplexed,choi2010entanglement,saglamyurek2011broadband,yang2016efficient,chou2007functional}. As the state of the art, up to four atomic ensemble quantum interfaces have been entangled through the heralded photon detection \cite{choi2010entanglement}.

In this paper, we report a significant advance in this direction by
experimentally generating multipartite entanglement between $25$, $16$, and $%
9$ individually addressable atomic quantum interfaces, and confirm genuine $22$, $%
14 $, and $9$ partite entanglement respectively for these cases with a high
confidence level by measuring the entanglement witness. Through programmable
control and heralded detection of photon interference from a two-dimensional
array of micro atomic ensembles, we generate and experimentally confirm the
multipartite W-state entanglement, which is one of the most robust types of
many-body entanglement and has applications in various quantum information
protocols \cite{Wstate,WT,IonW,haas2014entangled,mcconnell2015entanglement}. Tens to thousands of atoms in a single
atomic ensemble have been entangled with a heralded photon detection \cite%
{haas2014entangled,mcconnell2015entanglement}. In those cases, however, the
atoms are not separable or individually addressable and we do not have
multipartite entanglement between individual quantum interfaces. In other
experimental systems, up to $14$ ions \cite{monz201114}, $10$ photons \cite%
{wang2016experimental}, and $10$ superconducting qubits \cite{song201710}
have been prepared into genuinely entangled states. Those experiments
generate multipartite entanglement between individual particles, but each
particle alone cannot act as an efficient quantum interface to couple the memory
qubits with the flying photons. Our experiment achieves multipartite
entanglement between a record-high number of individually addressable
quantum interfaces and demonstrates an important enabling step towards the realization of
quantum networks, long-distance quantum communication, and multipartite
quantum information processing \cite
{briegel1998quantum,duan2001long,kimble2008quantum,hammerer2010quantum,sangouard2011quantum,collins2007multiplexed,Wstate,WT}.

\begin{figure}[tbp]
\includegraphics[width=12cm]{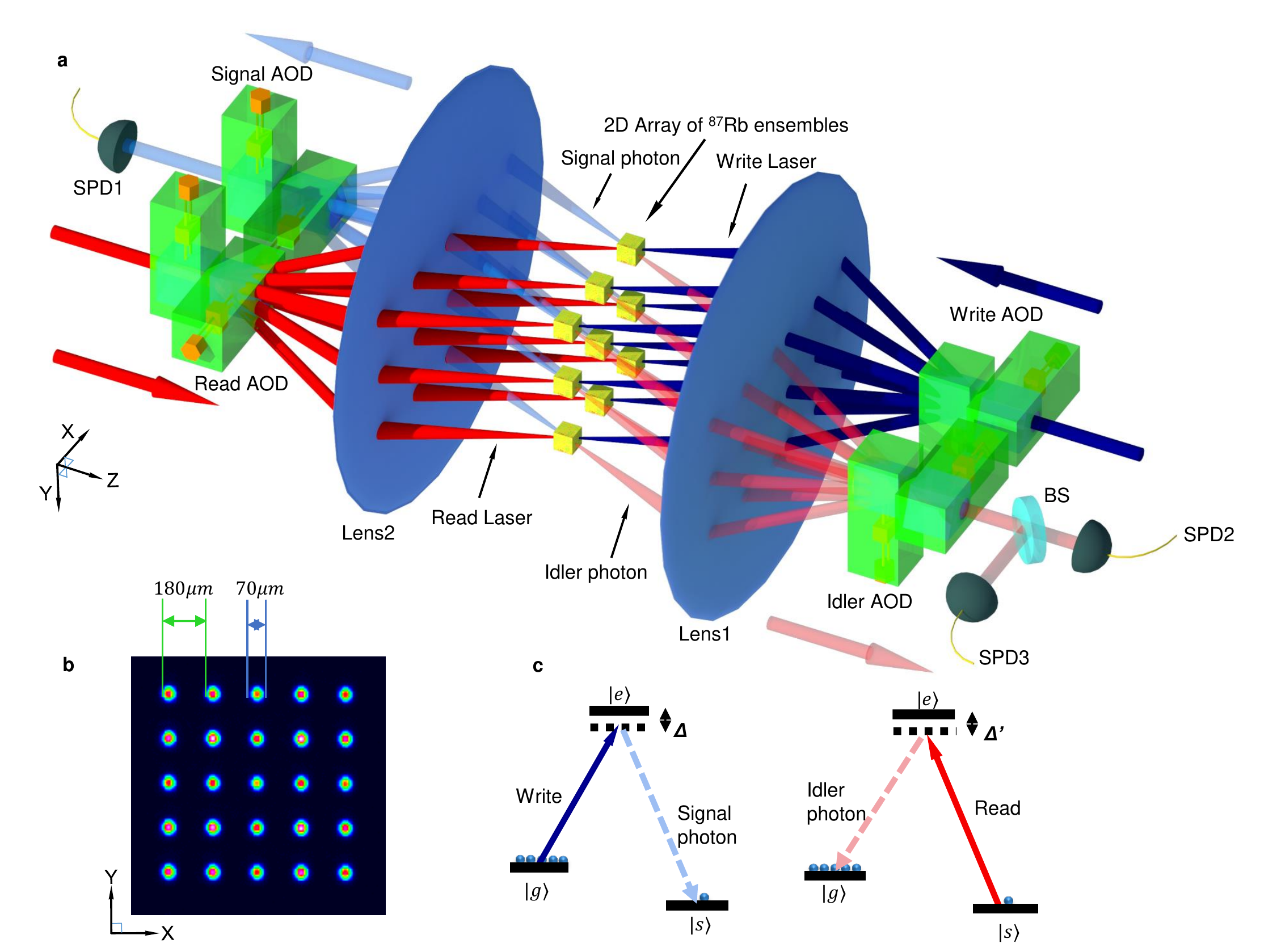}
\caption{\textbf{Experimental setup for generation and verification of multipartite entanglement
between a 2D array of atomic quantum interfaces.} \textbf{a}, We use a combination of the DLCZ scheme
and the programmable AOD multiplexer to generate multipartite entanglement of the W-state
type between the atomic spin waves in a 2D array of micro-ensembles. For clarity, we show a $
3\times3$ array, albeit we have also entangled $
4\times4$ and $5\times5$ ensemble arrays. The write laser beam is split coherently
into $9$ paths to simultaneously excite the $3\times3$ $ {}^{87}\mathrm{Rb}$ ensemble array
by the write AOD multiplexer which contains two
orthogonal deflectors placed in the X and Y
directions. The lens after the AOD multiplexer focuses the beams and at the same time maps different angles of the deflected
beams to different positions in a big atomic cloud forming individual micro-ensembles. The
scattered signal photon modes are combined phase coherently by the lens 2 and the signal AOD de-multiplexer, and then
coupled into a single-mode fiber with output detected by the single photon detector
(SPD1). To verify multipartite entanglement, we use a programmable AOD multiplexer
and de-multiplexer in the paths of the read beam and the idler photon mode to detect the atomic
spin waves from different micro-ensembles in several complementary bases. To bound the double
excitation probability, the idler photon mode is split by a 50/50
beam-splitter (BS) and detected by two single-photon detectors (SPD2 and SPD3) for registration of the
three-photon coincidence (together with the SPD1).
\textbf{b,} Illustration of the $5\times5$ array from multiplexing of a laser beam at the position of the atomic ensemble.
This image is obtained by shining a laser beam into the signal
single-mode fiber which is multiplexed by the signal AOD and captured by a CCD camera at
the position of atomic ensemble. The separation between adjacent signal modes is $180\,\mu$m both
in X and Y directions, and the Gaussian diameter of both the signal and the idler modes is
$70\,\mu$m. \textbf{c,} Relevant atomic energy levels and their couplings to the write/read
laser beams and the signal/idler photon modes,
with $|g\rangle\equiv|5S_{1/2}, F=2\rangle$, $|s\rangle\equiv|5S_{1/2},F=1\rangle
$, and $|e\rangle\equiv|5P_{1/2},F'=2\rangle$. The write (read) laser beam
is red detuned at $\Delta =10\,$MHz $(\Delta ^{\prime }=0)$, respectively, at the center micro-ensemble.}
\end{figure}

\section{Results}
\subsection{Experimental setup}

Our experimental setup is illustrated in Fig.~1. We divide a macroscopic $%
^{87}$Rb atomic ensemble into a two-dimensional array of micro ensembles
\cite{pu2017experimental}. Each micro-ensemble is optically dense and thus
can serve as an efficient quantum interface. Different micro-ensembles can
be individually or collectively accessed in a programmable way through
electric control of a set of cross-placed acoustic optical deflectors (AODs)
\cite{lan2009multiplexed,pu2017experimental}, with details described in the Methods
section. Programmable control of the experimental setup plays an
important role for scalable generation of entanglement \cite{Prog}.

We use a variation of the Duan-Lukin-Cirac-Zoller (DLCZ) scheme to generate
multipartite entanglement between the two-dimensional array of micro atomic
ensembles \cite{duan2001long}. The information in each atom is carried by the hyperfine levels $%
|g\rangle \equiv |5S_{1/2},F=2\rangle $ and $|s\rangle \equiv
|5S_{1/2},F=1\rangle $ in the ground-state manifold. All the atoms are
initially prepared to the state $|g\rangle $ through optical pumping, and
this initial state is denoted as $|0\rangle $ for each micro-ensemble.
Through the DLCZ scheme, a weak write laser pulse can induce a Raman
transition from $|g\rangle $ to $|s\rangle $, scatter a photon to the signal
mode in the forward direction with an angle of $2^{\circ }$ from the write
pulse, and excite a single atom into the corresponding collective spin-wave
mode. This state with one collective spin-wave excitation is denoted as $%
|1_{i}\rangle $ for the $i$th micro-ensemble.

We generate multipartite entanglement of the W-state type between
micro-ensemble quantum interfaces \cite{choi2010entanglement,Wstate,WT,IonW,haas2014entangled,mcconnell2015entanglement}. For $N$ micro-ensembles, an ideal W
state has the form%
\begin{equation}
|W_{N}\rangle =\frac{1}{\sqrt{N}}\sum_{i=1}^{N}e^{i\phi _{i}}|00\dots
1_{i}\dots 00\rangle ,
\end{equation}
where for the $i$th component we have a stable but adjustable phase factor $e^{i\phi _{i}}$ and a
single collective spin-wave excitation in the $i$th micro-ensemble. The W state
corresponds to a type of extremal multipartite entangled state most robust to
the particle loss \cite{Wstate} and has applications in implementation of quantum information
protocols \cite{duan2001long,kimble2008quantum,Wstate,WT,IonW,haas2014entangled,mcconnell2015entanglement}. To generate the W state entanglement between $N$
micro-ensembles, we split the write laser pulse into $N$ beams by the write
AODs as shown in Fig.~1, and coherently combine the signal photon modes from
$N$ micro-ensembles by the signal AODs with equal weight into a single
direction which is coupled to a single-mode fiber for detection. When we
register a signal photon by the detector, this photon is equally likely to
come from each micro-ensemble, which has an atomic excitation in the
corresponding spin-wave mode. The final state of $N$ micro-ensembles is
described by the W state (1) in the ideal case as the AODs maintain
coherence between different optical superposition paths.

\subsection{Verification of multipartite entanglement }

The experimentally prepared state differs from the ideal form (1) from
contribution of several noises and imperfections. First, there is a small but
nonzero probability to generate double or higher-order excitations of the photon-spin-wave
pair. Second, the spin-wave mode could be in the vacuum state when we registered
a photon due to the imperfect atom-photon correlation or the excitation loss in the atomic memory. Finally, even with
exactly one spin-wave excitation, it may not distribute equally or
perfectly coherently among $N$ micro-ensembles. The experimental state $\rho
_{e}$ can be expressed as
\begin{equation}
\rho _{e}=p_{0}\rho _{0}+p_{1}\rho _{1}+p_{2}\rho _{2},
\end{equation}%
where $p_{0},p_{1},p_{2}$ and $\rho _{0},\rho _{1},\rho _{2}$ denote
respectively the population and the corresponding density matrix with zero,
one, and double excitations in the spin-wave modes. The state fidelity is
defined as $F=\langle W_{N}|\rho _{e}|W_{N}\rangle =p_{1}\langle W_{N}|\rho
_{1}|W_{N}\rangle $. In the above equation (2), we have cut the expansion to the
second order excitations by neglecting tiny higher-order terms. If we assume a Poisson
distribution for the number of excitations (which is the case for a parametric light-atom interaction under weak
pumping), we can estimate the contribution of the higher order excitations from the measured
$p_2/p_1$. Their influence turns out to be negligible to all our following results (see the supplementary materials S2 \cite{sm}).

\begin{figure}[ptb]
\includegraphics[width=12cm]{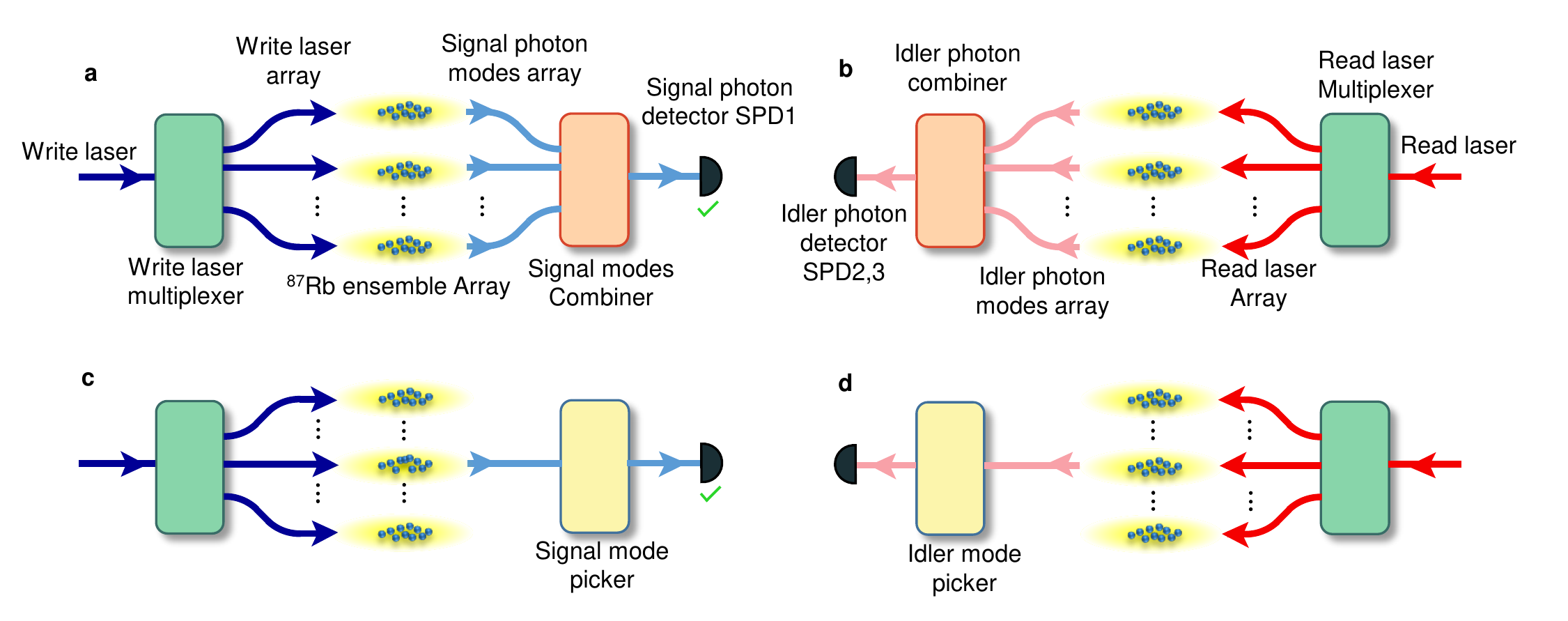}
\caption{ \textbf{Programmable coupling configurations for entanglement generation and verification.}
\textbf{a,} The coupling configuration to generate multipartite entanglement, where the write AODs
split the optical paths and the signal AODs coherently combine the paths. \textbf{b,} The detection configuration
for the measurement of fidelity, where the read AODs deliver the read beams to all the micro-ensembles to transfer
the atomic spin-wave excitations to idler photons and the idler AODs combine coherently the idler modes from different ensembles with equal weight for detection in the superposition basis. \textbf{c,d} The write and the read AODs in c and d are configured in the same way as those in a and b,
but the signal and the idler AODs are programmed to successively detect the signal/idler photon from each individual micro-ensemble. The configurations c and d combined are used to calibrate the retrieval efficiency for each micro-ensemble, and the
configurations a and d combined are used to detect the excitation population in each ensemble after the W state preparation (see the supplementary materials S2 for details \cite{sm}).   }
\end{figure}

\begin{figure}[ptb]
\includegraphics[width=8cm]{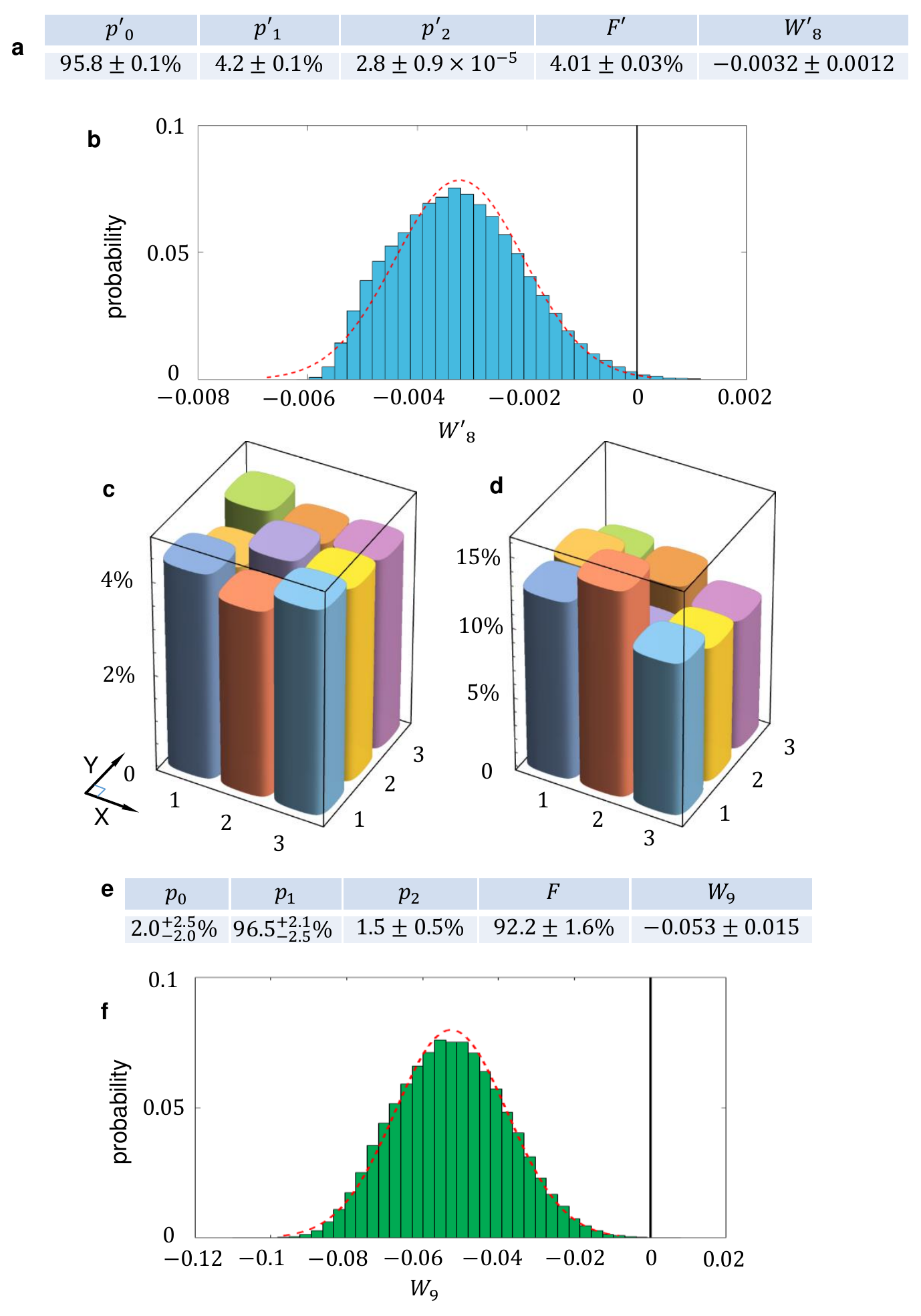}
\caption{ \textbf{Entanglement verification for the $3 \times 3 $ array of atomic ensembles.} \textbf{a,} The measured values, together with the $68\%$ confidence intervals (corresponding to the region within one standard deviation if the distribution is Gaussian), for the population $p'_0, p'_1, p'_2$, the fidelity $F'$, and the entanglement witness $W'_8$ for the idler photon modes that are directly measured. The entanglement in the retrieved idler photon modes provides a lower bound to the entanglement in the collective spin-wave modes in different atomic ensembles. The optimized parameters in the witness $W'_8$ are given by $\alpha'_8 = 2.259\times 10^{-3}$, $\beta'_8=0.7898$, and $\gamma'_8 = 49.13$. \textbf{b,} The distribution of entanglement witness $W'_8$,
where $W'_8<0$ implies 8-partite genuine entanglement. The probability with $W'_8<0$ is $99.5\%$ from this measurement. \textbf{c,} The measured retrieval efficiency for each of the $3 \times 3 $ atomic ensemble array. \textbf{d,} The measured spin-wave excitation population in each of the $3 \times 3 $ atomic ensemble array after the W state preparation. \textbf{e,} The measured values, together with the $68\%$ confidence intervals for the population $p_0, p_1, p_2$, the fidelity $F$, and the entanglement witness $W_9$ for the collective spin-wave modes in different atomic ensembles after correction of the retrieval efficiency through the above measurements. The optimized parameters in the witness $W_9$ are given by $\alpha_9 = 0.369$, $\beta_9=0.889$, and $\protect%
\gamma_9 = 0.268$. \textbf{f,} The distribution of entanglement witness $W_9$,
where $W_9<0$ implies 9-partite genuine entanglement. The probability with $W_9<0$ is $99.98\%$ from this measurement.}
\end{figure}

\begin{figure}[ptb]
\includegraphics[width=8cm]{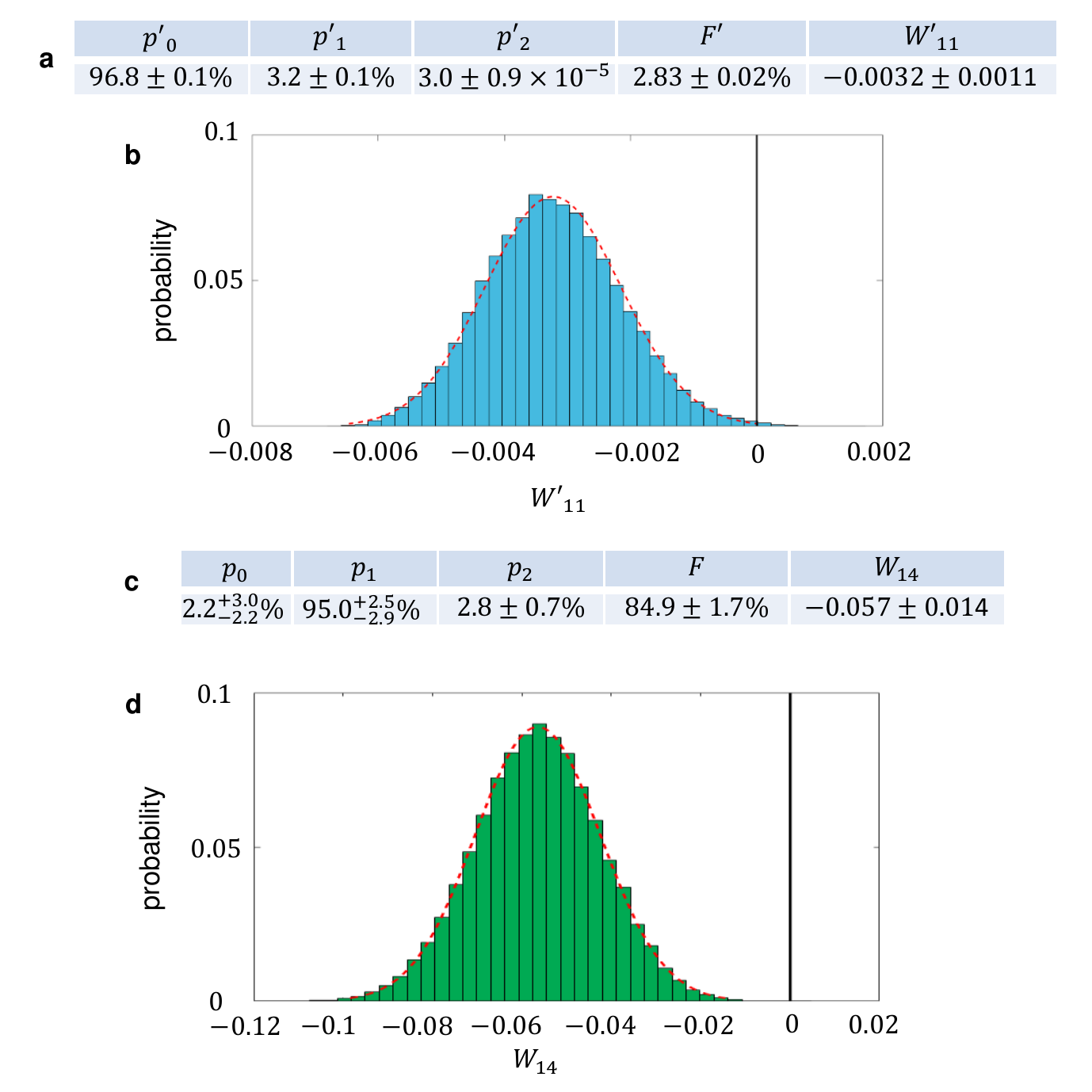}
\caption{\textbf{Entanglement verification for the $4 \times 4 $ array of atomic ensembles.} \textbf{a,} The measured values, together with the $68\%$ confidence intervals, for the population $p'_0, p'_1, p'_2$, the fidelity $F'$, and the entanglement witness $W'_{11}$, for the directly measured idler photon modes retrived from the $4 \times 4 $ atomic ensemble array. The optimized parameters in the witness $W'_{11}$ are given by $\alpha'_{11} = 3.152\times10^{-3} $, $\beta'_{11}=0.6370$, $\gamma'_{11} = 58.14$. \textbf{b,} The distribution of entanglement witness $W'_{11}$ for the $4 \times 4 $ idler photon modes. The probability with $W'_{11}<0$ is $99.7\%$ from these measurements. \textbf{c,} The measured values, together with the $68\%$ confidence intervals, for the population $p_0, p_1, p_2$, the fidelity $F$, and the entanglement witness $W_{14}$, for the $4 \times 4 $ atomic ensemble array after correction of the retrieval efficiency. The optimized parameters in the witness $W_{14}$ are given by $\alpha_{14} = 0.635 $, $\beta_{14}=0.813$, $\gamma_{14} = 0.240$. \textbf{d,} The distribution of entanglement witness $W_{14}$ for the $4 \times 4 $ case. The probability with $W_{14}<0$ is $99.997\%$ from these measurements.}
\end{figure}

\begin{figure}[ptb]
\includegraphics[width=8cm]{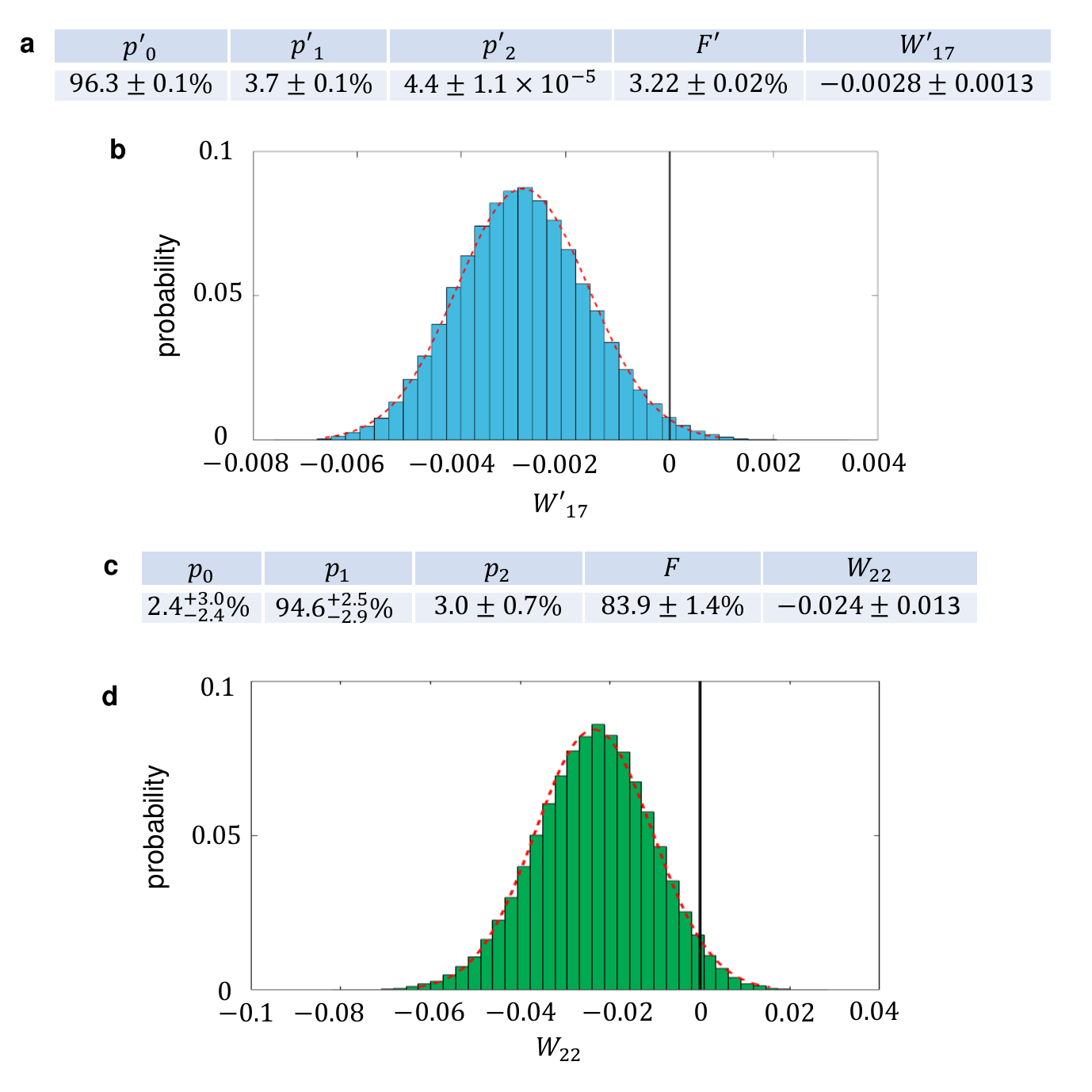}
\caption{\textbf{Entanglement verification for the $5 \times 5 $ array of atomic ensembles.} \textbf{a,} The measured values, together with the $68\%$ confidence intervals, for the population $p'_0, p'_1, p'_2$, the fidelity $F'$, and the entanglement witness $W'_{17}$, for the directly measured idler photon modes retrieved from the $5 \times 5 $ atomic ensemble array. The optimized parameters in the witness $W'_{17}$ are given by $\alpha'_{17} = 3.317\times10^{-3} $, $\beta'_{17}=0.6516$, $\gamma'_{17} = 53.65$. \textbf{b,} The distribution of entanglement witness $W'_{17}$ for the $5 \times 5 $ idler photon modes. The probability with $W'_{17}<0$ is $98.4\%$ from these measurements. \textbf{c,} The measured values, together with the $68\%$ confidence intervals, for the population $p_0, p_1, p_2$, the fidelity $F$, and the entanglement witness $W_{22}$, for the $5 \times 5 $ atomic ensemble array after correction of the retrieval efficiency. The optimized parameters in the witness $W_{22}$ are given by $\alpha_{22} = 0.550 $, $\beta_{22}=0.840$, $\gamma_{22} = 0.244$. \textbf{d,} The distribution of entanglement witness $W_{22}$ for the $5 \times 5 $ case. The probability with $W_{22}<0$ is $96.5\%$ from these measurements.}
\end{figure}

To verify multipartite quantum entanglement between $N$ quantum
interfaces, we use entanglement witness to
lower bound the entanglement depth $k$ ($k\leq N$) \cite{EDepth}, which means the state $\rho _{e}$ has at least $k$-partite
genuine quantum entanglement \cite{guhne2009entanglement}. An entanglement witness appropriate for the
W-type entangled state is given by $\mathcal{W}_{k}=\alpha _{k}P_{0}+\beta
_{k}P_{1}+\gamma _{k}P_{2}-|W_{N}\rangle \langle W_{N}|$ \cite%
{guhne2009entanglement}, where $P_{n}$ ($n=0,1,2$) denote the projectors
onto the subspace with $n$ excitations in the spin-wave modes and the
parameters $\alpha _{k},\beta _{k},\gamma _{k}\geq 0$ are numerically
optimized (see the supplementary materials S1 \cite{sm}) such that for any state $\rho _{a}$ with entanglement depth
less than $k$, the witness is non-negative, i.e. $\mathrm{tr}[\mathcal{W}%
_{k}\rho _{a}]=\alpha _{k}p_{0}+\beta _{k}p_{1}+\gamma _{k}p_{2}-F\geq 0$.
Therefore, $\mathrm{tr}[\mathcal{W}_{k}\rho _{e}]<0$ serves as a sufficient
condition to verify that we have at least $k$-partite genuine entanglement
among the $N$ quantum interfaces.
Note that this witness does not require
$p_{0}+p_{1}+p_{2}=1$, so it also applies in the case with $p_{0}+p_{1}+p_{2}<1$ when we
consider small higher-order excitations, although the corrections turn out to be
negligible for all our following results \cite{sm}).

To bound the entanglement depth, we experimentally measure the fidelity $F$
and the population $p_{0},p_{1},p_{2}$. The detailed measurement procedure
is explained in the supplementary materials S2 \cite{sm}. The spin-wave excitation in
each quantum interface is retrieved to the idler photon for detection by a
read laser beam. Our measurement is directly on the state of the retrieved photon,
which can be represented by a form similar to equation (2) for the spin-wave modes.
Due to the limited retrieval efficiency, detector inefficiency, and the associated photon loss, the
detected idler photon modes have much larger vacuum components, and their corresponding
parameters are denoted as $F'$ and $p'_{0},p'_{1},p'_{2}$.
Because this retrieval process is a local operation,
the entanglement in the retrieved photonic
modes provides a lower bound to the entanglement in the collective spin-wave modes in the atomic
ensembles \cite{choi2010entanglement}.

The fidelity $F'$ and the populations $p'_0,p'_1,p'_2$ of the idler photon are determined in the following way.
We first measure the double excitation probability $p'_{2}$ from the photon intensity correlation
of the two single photon detectors in the idler modes, conditioned on
a photon click in the signal mode. Then $p'_{1},p'_{0}$ and
$F'$ are measured by programming the four sets (write, signal, read, and idler) of
AODs in different configurations as shown in Fig. 2 (see details in the supplementary materials S2 and figures S1-S4 there \cite{sm}). When we
measure the population $p'_{1}$, the idler AOD successively picks up the
output photon mode of each individual micro-ensemble for detection; and as for the fidelity $F'$, the idler AOD coherently combines the output idler modes from the $N$ micro-ensembles with equal
weight to the single-mode
fiber for detection, which gives an effective projection to the state $%
|W_{N}\rangle $.
Note that the fidelity measurement is sensitive to the relative phase information between different idler photon modes as these modes interfere at the AODs through the coherent combination.
After $F'$ and $p'_0,p'_1,p'_2$ are measured, we calibrate
the retrieval efficiency for each micro-ensemble, and finally derive the fidelity
$F$ and populations $p_0,p_1,p_2$ of the spin-wave modes from the measured idler photon statistics \cite{Chou2004}.
The detailed conversion procedure is described in the supplementary materials S2 \cite{sm}.

We have performed the entanglement preparation and verification experiments
with $3\times 3$, $4\times 4$ and $5\times 5$ arrays of micro-ensembles.
For $9$ individually addressable micro-ensembles, the results are shown in
Fig.~3. We present the parameters for the idler photon state in Fig.~3a, and the
probability to have $8$-partite entanglement is $99.5\%$ for the photon state. After conversion with the calibrated retrieval efficiency, we find that the state of the atomic micro-ensembles has a high fidelity of $F=(92.2\pm 1.6)\%$ to be in the $9$-partite W state. In Fig.~3d, we show the distribution of the entanglement witness $W_{9}=\mathrm{tr}[\mathcal{W}%
_{9}\rho _{e}]$ from the experimental data. From this
distribution, we conclude with a confidence level of $99.98\%$ that we have
generated genuine $9$-partite quantum entanglement among the $9$ atomic ensembles.

In Fig.~4 and Fig.~5, we show the experimental results for $16$
and $25$ micro-ensembles. In these cases,
the fidelity is not high enough to prove all of them are genuinely
entangled. The calibrated fidelities $F$ for the atomic states are $(84.9\pm 1.7)\%$ and $%
(83.9\pm 1.4)\%$, respectively. With more ensembles, it becomes harder to maintain the
uniformity in the optical depth and the laser excitation probability for
each ensemble, which causes the fidelity to decay. However, we can still use the entanglement witness to demonstrate a high
entanglement depth. As shown in Fig.~4 and Fig.~5,
for $16$ ensembles, we have confirmed $11$-partite
entanglement in the retrieved idler photon modes with a confidence level of $99.7\%$ and $14$-partite entanglement
between the spin-wave modes in the $16$ micro-ensembles with a confidence level of
 $99.997\%$ after correction with the calibrated retrieval efficiency; and for $25$
ensembles, we have confirmed $17$-partite entanglement in the retrieved photonic modes with
a confidence level of $98.4\%$ and $22$-partite entanglement between the $25$ micro-ensembles with a confidence level of $%
96.5\%$.

\section{Discussion}

Our experimental preparation of multipartite entanglement in a
record-high number of individually addressable quantum interfaces represents a
significant milestone in quantum state engineering. Through programming of
AODs to control intrinsically stable optical interference paths, the
entanglement preparation and verification techniques developed in this
experiment are fully scalable to a larger number of quantum interfaces.
It is feasible to use AODs to program and direct the focused laser beams to
hundreds of micro-ensembles \cite{pu2017experimental}. The number of entangled
ensembles in our current experiment is basically limited by the size of the whole
atomic cloud and the available optical depth. With the use of double magneto-optical traps
for more efficient atom loading, we can significantly increase the size of the atomic cloud,
the optical depth, and the retrieval efficiency for the stored photons. In that case,
we should be able to get hundreds of micro-ensembles entangled by the same control setup
and entanglement verification techniques reported in this experiment.
Generation of multipartite entanglement between many individually
addressable quantum interfaces demonstrates an important step towards the realization
of quantum networks \cite{kimble2008quantum,duan2001long}, long-distance quantum communication \cite{duan2001long,briegel1998quantum,sangouard2011quantum}, and
multipartite quantum information processing \cite{kimble2008quantum,choi2010entanglement,Wstate,WT}.

\textbf{Note Added.} After post of this work to arxiv (arXiv:1707.09701), we became aware of related independent works
in Refs.~\cite{Simon2017,Gisin2017}, which report generation of multi-particle W state entanglement in solid-state ensembles. Compared with those experiments, we realized multipartite entanglement between spatially separated micro-ensembles of neutral atoms which are individually
accessible by focused laser beams with programmable control of the AODs. We thank C. Simon for bringing Refs.~\cite{Simon2017,Gisin2017} to our attention.

\section*{\protect\large Materials and Methods}

\textbf{Experimental methods}. A ${}^{87}$Rb atomic cloud is loaded into a
magneto-optical trap (MOT). For cooling and trapping of the atoms in the
MOT, a strong cooling beam, red detuned to the D2 cycling transition $%
|g\rangle \equiv |5S_{1/2},F=2\rangle \rightarrow |5P_{3/2},F=3\rangle $ by $%
12\,$MHz, is used. The repumping laser, resonant to the $|s\rangle \equiv
|5S_{1/2},F=1\rangle \rightarrow |5P_{3/2},F=2\rangle $ transition, pumps
back those atoms which fall out of the cooling transition. The temperature
of the atoms is about $300\,\mu $K in the MOT. The atoms are then further
cooled by polarization gradient cooling (PGC) for $1\,$ms. The PGC is
implemented by increasing the red detuning of the cooling laser to $60\,$%
MHz, and reducing the intensity to half of the value at the MOT loading
stage. At the same time, the repumping laser intensity is decreased to $%
0.5\% $ of the value at the loading phase, and the magnetic gradient coil is
shut off. The temperature is reduced to about $30\,\mu $K after this process
and the size of the MOT remains almost the same. After the PGC some
atoms are scattered to the $|s\rangle $ state, and we use a $100\,\mu
$s repumping pulse to pump all the atoms back to $|g\rangle $. During the
storage, the ambient magnetic field is not compensated, so the retrieval
efficiency of the collective spin-wave excitation undergoes Larmor
precession. In our case, the Larmor period is $5.8\,\mu $s. The time
interval between the read and the write pulses is set to this Larmor period
to achieve the highest retrieval efficiency for the idler photon.

The experimental sequence begins with a write pulse of $100\,$ns long, which
is split by the write AODs to $N$ paths to excite the two-dimensional (2D)
array of atomic ensembles. If no signal photon is detected, a clearance
pulse identical to the read pulse will pump the atoms back to $|g\rangle $.
The write-clearance sequence is repeated until a signal photon is detected.
Upon detection of the signal photon, the corresponding collective spin-wave
excitation is stored in the atomic ensemble for a controllable period of
time and then retrieved by a read pulse to a photon in the idler mode. The
conditional control of write/read pulses is implemented by a
field-programmable gate array (FPGA). The signal or idler photons collected
by the single-mode optical fiber are directed to a single-photon counting
module (SPCM). The photon countings and their coincidence are registered
through the FPGA.\newline

\textbf{Control of acoustic optical deflectors}. The radio-frequency (RF)
signal is generated by two $4$-channel arbitrary-waveform generators (AWG,
Tektronix 5014C). One of the AWG supplies the RF for write, read, signal,
and idler acoustic optical deflectors (AODs, AA DTSXY-400) in the $X$
direction, and the other supplies the RF for the AODs in the $Y$ direction.
The outputs of the AWG channels are amplified by a $2\,$W RF amplifier
(Mini-circuits, ZHL-1-2W) to drive the AODs.

The nonlinearity in the amplifier and the AODs could induce other unwanted
frequency components, which cause imperfections in the mode multiplexing and
de-multiplexing. By carefully tuning the relative phases in read, signal,
and idler AODs as discussed in \cite{endres2016atom}, we can attenuate the
influence from these unwanted frequency components by an extinction ratio
about $120\,$dB, which becomes negligible for our experiment.

Although the AODs split the optical paths into many different branches, the
relative optical phases between different branches are intrinsically stable
as different optical paths in our experiment go through the same optics
elements. This is an important advantage which eliminates the need of
complicated active phase stabilization for many optical interferometer loops
in our experiment. The relative phases between different superposition paths
are adjusted in experiments by controlling the phases of different RF
frequency components that drive the write AODs. \newline

\textbf{Acknowledgements} We thank A. Kuzmich and Y.-M. Liu
for discussions. This work was supported by the Ministry of Education of China
and the Tsinghua-QTEC Joint Lab on quantum networks.
LMD acknowledges in addition support from
the ARL CDQI program.

\textbf{Author Contributions} L.M.D. conceived the experiment and supervised
the project. Y.F.P., N.J., W.C., C.L., S.Z. carried out the experiment. Y.K.W.
optimized the entanglement witness. L.M.D., Y.F.P., Y.K.W. wrote the manuscript.

\textbf{Author Information} The authors declare no competing financial
interests. Correspondence and requests for materials should be addressed to
L.M.D. (lmduan@umich.edu).

\textbf{Data and materials availability:} All data needed to evaluate the conclusions in the paper are present in the paper
and/or the Supplementary Materials. Additional data related to this paper may be requested from the authors.

\newpage

\section*{\protect\large Supplementary text}

\section{ Section S1. Entanglement witness for W-type states}

Ideally, we should generate the W-type multipartite entangled states. Due to
noise and imperfections, the experimentally prepared state is always mixed.
To verify multipartite entanglement in the proximity of the W states, we use
the following entanglement witness introduced in Ref.~\cite{guhne2009entanglement}
\begin{equation}
\mathcal{W}_{k}=\alpha _{k}P_{0}+\beta _{k}P_{1}+\gamma
_{k}P_{2}-|W_{N}\rangle \langle W_{N}|,
\end{equation}%
where $P_{n}$ ($n=0,1,2$) is the projector onto the subspace with $n$
excitations ($n$ ($n\leq N$) qubits in the $\left\vert 1\right\rangle $
state), and
\begin{equation}
|W_{N}\rangle =\frac{1}{\sqrt{N}}\left( |10\cdots 00\rangle +|01\cdots
00\rangle +\dots +|00\cdots 01\rangle \right) \label{eq:2}
\end{equation}%
denotes the $N$-qubit W state, where we have neglected the unimportant
relative phases between the superposition terms as they can be absorbed into
the definition of the basis states. The parameters $\alpha _{k},\beta
_{k},\gamma _{k}\geq 0$ are chosen such that for any state $\rho $ with
entanglement depth $E_{d}$ less than $k$ (states without genuine $k$-partite
entanglement), the witness is non-negative, i.e., $\mathrm{tr}(\mathcal{W}%
_{k}\rho )\geq 0$.

In the following, we briefly describe how to optimize this entanglement witness following the
derivation given in Ref. \cite{Guhne2009}, which is referred to for the full details of the arguments.
We pay particular attention to optimizing the parameters $\alpha _{k},\beta _{k},\gamma
_{k}$ for our experimental configurations. Since a general density operator $\rho $ can always be
expressed as a convex combination of pure states, it suffices to consider
the non-negativity over pure states $|\phi \rangle $ with $E_{d}<k$.
Furthermore, due to the permutation symmetry of the W state and the $P_{n}$
operators, we can write $|\phi \rangle =|a\rangle _{1,\cdots ,l}|b\rangle
_{l+1,\cdots ,N}$ ($l,N-l\leq k$) without loss of generality. Here we are
mainly interested in the case where $k$ is close to $N$ and hence we assume $%
k\geq 2N/3$. The above expression also includes the case where $|\phi
\rangle $ can be separated into the tensor product of more than two parts.
If we find such an entanglement witness $\mathcal{W}_{k}$ characterized by
the parameters $\alpha _{k}$, $\beta _{k}$, $\gamma _{k}$, and if the
experimentally generated state $\rho _{e}$ satisfies $\mathrm{tr}(\mathcal{W}%
_{k}\rho _{e})=\alpha _{k}p_{0}+\beta _{k}p_{1}+\gamma _{k}p_{2}-F<0$, we
can conclude that the state $\rho _{e}$ must possess at least genuine $k$%
-partite entanglement. The parameters $p_{0},p_{1},p_{2}$ in the above
witness denote the population with zero, one, or double excitations in the
spin-wave modes and $F\equiv \langle W_{N}|\rho _{e}|W_{N}\rangle $ denotes
the state fidelity. The parameters $p_{0},p_{1},p_{2},F$ are directly
measured in our experiment.

The component state $|a\rangle _{1,\cdots ,l}$ (and similarly $|b\rangle
_{l+1,\cdots ,N}$) can be generally expanded as
\begin{equation}
|a\rangle _{1,\cdots ,l}=a_{0}|g\rangle _{1,\cdots ,l}+a_{1}|e_{1}\rangle
_{1,\cdots ,l}+\cdots +a_{l}|e_{l}\rangle _{1,\cdots ,l}
\end{equation}%
where $|g\rangle _{1,\cdots ,l}=|00\cdots 0\rangle _{1,\cdots ,l}$ denotes
the ground state with all the qubits in the $|0\rangle $ state, $%
|e_{1}\rangle _{1,\cdots ,l}\propto P_{1}|a\rangle _{1,\cdots ,l}$ is a
normalized state with exactly one excitation, and $|e_{l}\rangle _{1,\cdots
,l}$ denotes a state with exactly $l$ excitations. Our purpose is to find out
the optimal parameters $\alpha _{k},\beta _{k},\gamma _{k}$ so that for any
state $|\phi \rangle $ with above decomposition, we have $\mathrm{tr}(%
\mathcal{W}_{k}|\phi \rangle \langle \phi |)\geq 0$. The non-negativity of
the witness is not affected by normalization of the state. Suppose we have a
state $|\phi \rangle =(a_{0}|g\rangle +a_{1}|e_{1}\rangle )_{1,\cdots
,l}(b_{0}|g^{\prime }\rangle +b_{1}|e_{1}^{\prime }\rangle )_{l+1,\cdots ,N}$
whose witness is non-negative, i.e., $\mathrm{tr}(\mathcal{W}_{k}|\phi
\rangle \langle \phi |)\geq 0$. Now if we keep $a_{0}$, $a_{1}$, $b_{0}$ and
$b_{1}$ unchanged but introduce non-zero $a_{2},\cdots ,a_{l},b_{2},\cdots
,b_{N-l}$ terms, the projection onto $P_{0}$, $P_{1}$ and $|W_{N}\rangle
\langle W_{N}|$ remain unaffected while the projection on $P_{2}$ may
increase, because the added terms have at least two excitations. Therefore
this new state is guaranteed to have a non-negative witness. In other words,
to test the non-negativity of the entanglement witness, we only need to
consider bi-decomposable pure states $|\phi \rangle $ with each part staying
in the subspace of no more than one excitation. For the same reason, only
the completely symmetric state $|W_{l}\rangle $ ($|W_{N-l}\rangle $) needs
to be considered in the one-excitation subspace, since a one-excitation
state orthogonal to the symmetric state is also orthogonal to $|W_{N}\rangle
$ but still contributes to the $P_{1}$ and $P_{2}$ terms.

Through the above reasoning, we only need to find optimal $\alpha _{k},\beta
_{k},\gamma _{k}$ such that for any $l,N-l\leq k$ and any complex numbers $%
a_{0},a_{1},b_{0},b_{1}$ with $%
|a_{0}|^{2}+|a_{1}|^{2}=|b_{0}|^{2}+|b_{1}|^{2}=1$, the state
\begin{equation}
|\phi \rangle =(a_{0}|g\rangle +a_{1}|W_{l}\rangle )_{1,\cdots
,l}(b_{0}|g^{\prime }\rangle +b_{1}|W_{N-l}\rangle )_{l+1,\cdots ,N}
\end{equation}%
has non-negative witness $\mathrm{tr}(\mathcal{W}_{k}|\phi \rangle \langle
\phi |)\equiv f\geq 0$, which can be expressed as
\begin{eqnarray}
f &=&\alpha _{k}|a_{0}|^{2}|b_{0}|^{2}+\beta
_{k}(|a_{0}|^{2}|b_{1}|^{2}+|a_{1}|^{2}|b_{0}|^{2})  \label{eq:polynomial} \\
&&+\gamma _{k}|a_{1}|^{2}|b_{1}|^{2}-\frac{1}{N}\left\vert \sqrt{N-l}%
a_{0}b_{1}+\sqrt{l}a_{1}b_{0}\right\vert ^{2}.  \notag
\end{eqnarray}%
The parameters $\alpha _{k},\beta _{k},\gamma _{k}$ should be chosen such
that the minimal value of $f$ is non-negative. Clearly this function is
minimized when $a_{0}$, $a_{1}$, $b_{0}$ and $b_{1}$ are in phase, so we can
choose $0\leq a_{0},a_{1},b_{0},b_{1}\leq 1$. Therefore, we can express them
as $a_{0}=\cos \theta _{1}$, $a_{1}=\sin \theta _{1}$, $b_{0}=\cos \theta
_{2}$, $b_{1}=\sin \theta _{2}$ ($0\leq \theta _{1},\theta _{2}\leq \pi /2$%
). With the new parameters $\theta _{1},\theta _{2}$, we have
\begin{align}
f=& \frac{1}{4}\bigg[\alpha _{k}(1+\cos 2\theta _{1})(1+\cos 2\theta
_{2})+2\beta _{k}(1-\cos 2\theta _{1}\cos 2\theta _{2})  \notag \\
& +\gamma _{k}(1-\cos 2\theta _{1})(1-\cos 2\theta _{2})-(1+\cos 2\theta
_{1})(1-\cos 2\theta _{2})  \notag \\
& +\frac{2l}{N}(\cos 2\theta _{1}-\cos 2\theta _{2})-\frac{2\sqrt{l(N-l)}}{N}%
\sin 2\theta _{1}\sin 2\theta _{2}\bigg].  \label{eq:target}
\end{align}

To find the minimum of this objective function, we calculate the partial
derivatives $\partial f/\partial \theta _{1}$ and $\partial f/\partial
\theta _{2}$ with respect to $\theta _{1}$ and $\theta _{2}$. Inside the
rectangular region $(0,\pi /2)\times (0,\pi /2)$, stationary points are
determined by $\partial f/\partial \theta _{1}=\partial f/\partial \theta
_{2}=0$, which gives,
\begin{align}
\tan 2\theta _{1}=& \frac{\frac{2\sqrt{l(N-l)}}{N}\sin 2\theta _{2}}{-\alpha
_{k}(1+\cos 2\theta _{2})+2\beta _{k}\cos 2\theta _{2}+\gamma _{k}(1-\cos
2\theta _{2})+(1-\cos 2\theta _{2})-\frac{2l}{N}},  \label{eq:iter_theta_1}
\\
\tan 2\theta _{2}=& \frac{\frac{2\sqrt{l(N-l)}}{N}\sin 2\theta _{1}}{-\alpha
_{k}(1+\cos 2\theta _{1})+2\beta _{k}\cos 2\theta _{1}+\gamma _{k}(1-\cos
2\theta _{1})-(1+\cos 2\theta _{1})+\frac{2l}{N}}.  \label{eq:iter_theta_2}
\end{align}%
To find the stationary point solution, we choose an arbitrary initial point
inside the region, say, with $\theta _{1}=\theta _{2}=\pi /4$, and then
apply the substitution Eq.~(\ref{eq:iter_theta_1}) and Eq.~(\ref%
{eq:iter_theta_2}) iteratively until the result converges. With this method,
we get the minimum of $f$ with respect to $\theta _{1},\theta _{2}$ for a
given $l$. This minimum of $f$ is also compared with the value of $f$ at the
boundary to get the absolute minimum of $f$ in the rectangular region.
Finally, the integer parameter $l$ is scanned so that we get the absolute
minimum of $f$, denoted as $f_{m}$, with respect to the parameters $l,\theta
_{1},\theta _{2}$.

We choose the parameters $\alpha _{k},\beta _{k},\gamma _{k}$ so that the
witness condition $f_{m}\geq 0$ is satisfied. There are infinite
combinations of $\alpha _{k},\beta _{k},\gamma _{k}$ that satisfy this
requirement. To find the optimal $\alpha _{k},\beta _{k},\gamma _{k}$ for
given experimental data $p_{0},p_{1},p_{2},F$, we choose a combination of $%
\alpha _{k},\beta _{k},\gamma _{k}$ that leads to the smallest (most
negative) entanglement witness $\mathrm{tr}(\mathcal{W}_{k}\rho _{e})=\alpha
_{k}p_{0}+\beta _{k}p_{1}+\gamma _{k}p_{2}-F$ because it is the negative
value of the witness $\mathrm{tr}(\mathcal{W}_{k}\rho _{e})$ that indicates
the existence of $k$-partite entanglement. The parameters $\alpha _{k},\beta
_{k},\gamma _{k}$ in the caption of Figs. 3 and 4 are determined in this way
for verification of genuine multipartite entanglement with $k=9,14,22$ and $%
N=9,16,25$, respectively. For measurement of the retrieved
idler photon modes, $\alpha' _{k},\beta'
_{k},\gamma' _{k}$ are optimized for the measured $p'_{0},p'_{1},p'_{2},F'$ in a similar way.

Note that although we assume a truncated number of excitations in the atomic micro-ensembles, as is shown
in Eq.~(2) in the main text, the entanglement witness we use here is exact. The above
derivation actually allows the existence of higher order excitations by simply taking
$p_0+p_1+p_2<1$. Later we will bound the effects of higher order excitations on the measured
probabilities and therefore give a lower-bound on the entanglement depth.

\section{Section S2. Experimental measurement of the entanglement witness}

To experimentally verify multipartite entanglement, we measure the
entanglement witness $\mathrm{tr}(\mathcal{W}_{k}\rho _{e})=\alpha
_{k}p_{0}+\beta _{k}p_{1}+\gamma _{k}p_{2}-F$, which reduces to measurements
of four parameters $p_{0},p_{1},p_{2},F$. To measure these parameters for
the spin-wave states in the atomic ensembles, all the detections are done
through the conversion of spin-wave excitations to the idler photons. First,
we need to calibrate the retrieval efficiency $\eta _{i}$ for each
micro-ensemble, which is defined as the probability to register a photon
count in the idler mode by the single-photon detector given a single
excitation in the corresponding collective spin-wave mode.

\begin{figure}[ptb]
\includegraphics[width=6in]{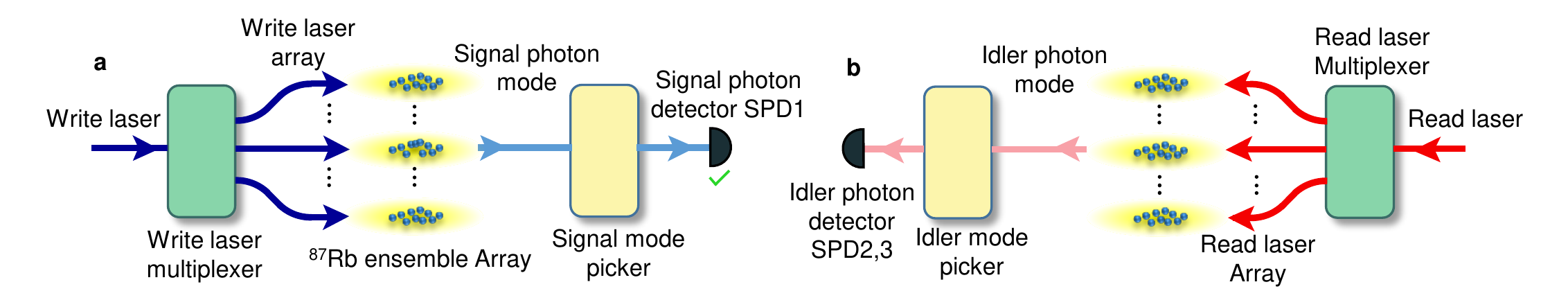}
\caption{ \textbf{Coupling configuration for the measurement of the retrieval efficiency of each micro-ensemble.}
\textbf{a,} The write AODs coherently split the write beam into $N$ paths
to excite all the micro-ensembles simultaneously, and the signal AODs are programmed
to detect the signal photon from one of the $N$ micro-ensembles. The detection of a signal photon by SPD1
heralds the generation of an atomic spin-wave excitation in the corresponding micro-ensemble.
\textbf{b,} The read AODs deliver the read beam to transfer all the atomic spin-wave excitations
to idler photons, and the idler AODs are programmed to detect the idler photon from the same
micro-ensemble where the signal photon is detected in \textbf{a}.
}
\end{figure}

\begin{figure}[ptb]
\includegraphics[width=6in]{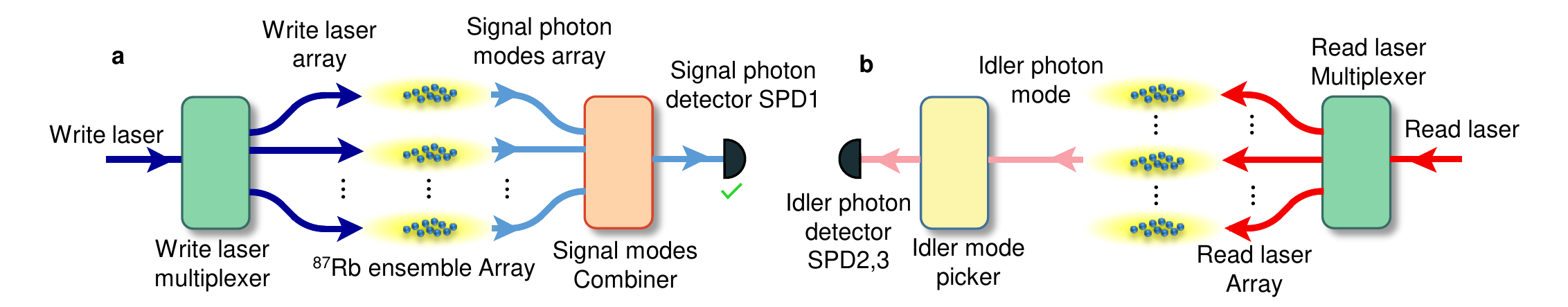}
\caption{ \textbf{Coupling configuration for the measurement of the excitation population of each micro-ensemble.}
\textbf{a,} The write AODs split
the optical paths and the signal AODs coherently combine the paths. The generation of a
spin-wave W state is heralded by the detection of a photon in the combined signal mode.
\textbf{b,} The read AODs coherently split the read beam to transfer all the atomic spin-wave excitations
to idler photons, and the idler AODs are programmed to detect the idler photon from one of the $N$
micro-ensembles to measure its excitation population.
}
\end{figure}

\begin{figure}[ptb]
\includegraphics[width=6in]{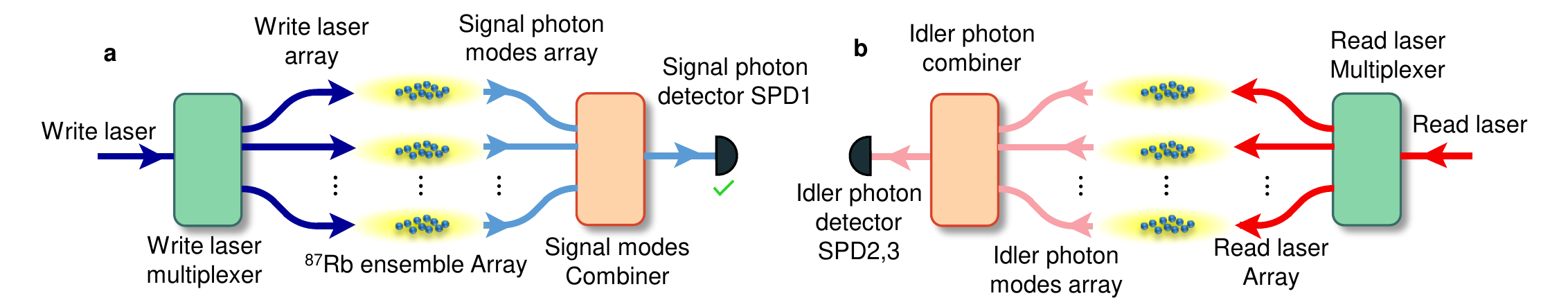}
\caption{ \textbf{Coupling configuration for the measurement of the $W$ state fidelity.}
\textbf{a,} The write AODs coherently split
the optical paths and the signal AODs coherently combine the paths. The generation of a
spin-wave W state is heralded by the detection of a photon in the combined signal mode.
\textbf{b,} The read AODs coherently split
the read beams to all the micro-ensembles to transfer the atomic spin-wave excitations to idler
photons and the idler AODs coherently combine the idler modes from different micro-ensembles
with equal weight for the detection in the superposition basis.
}
\end{figure}

We measure the retrieval efficiency by the setup shown in Fig.~S1.
Through control of the AODs, we successively excite and
measure each micro-ensemble through the standard DLCZ scheme. For the $i$th
ensemble, through the measured photon counts on the signal and idler modes
and their coincidence, we get the probability $P_{S}^{(i)}$ ($P_{I}^{(i)}$)
to record a photon count in the signal (idler) mode in each experimental
trial and the joint probability $P_{SI}^{(i)}$ to detect a coincidence. The
coincidence probability can be expressed as
\begin{equation}
P_{SI}^{(i)}=\eta _{i}P_{S}^{(i)}+P_{S}^{(i)}P_{I}^{(i)}, \label{eq:9}
\end{equation}%
where the second term $P_{S}^{(i)}P_{I}^{(i)}$ denotes the random
coincidence from two independent distributions and the first term denotes
the retrieved signal with the retrieval efficiency $\eta _{i}$. From the
above expression, we get $\eta _{i}=P_{SI}^{(i)}/P_{S}^{(i)}-P_{I}^{(i)}$,
which is inferred from the three measured qualities $%
P_{S}^{(i)},P_{I}^{(i)},P_{SI}^{(i)}$. For our experiment, the measured
retrieval efficiencies are close to $4\%$ for all the micro-ensembles. For
the $3\times 3$ micro-ensemble array, the results of the measured $\eta _{i}$
are shown in Fig. 3c of the main text.

After determination of the retrieval efficiency $\eta _{i}$, we can then
measure the population $p_{0},p_{1},p_{2}$ and the fidelity $F$. In our
experiment, the double excitation probability $p_{2}$ is quite small. To
illustrate the basic idea of detection method, first we look at a simple
case by neglecting the contribution of $p_{2}$\ (later we will go to the
more realistic case by determining the small but nonzero $p_{2}$). Without
the contribution of $p_{2}$, the experimental density matrix has the
simplified form $\rho _{e}=p_{0}\rho _{0}+p_{1}\rho _{1}$. To measure $p_{1}$%
, we use the setup shown in Fig.~S2. After
preparation of the state with the excitation configuration in Fig.~S2a, we
successively pick up the idler mode from each micro-ensemble to measure the
photon counts as shown in Fig.~S2b. The measured probability $q_{i}$ to
record a photon count from the $i$th idler mode in each experimental trial
is given by%
\begin{equation}
q_{i}=\eta _{i}p_{1}\langle i|\rho _{1}|i\rangle,  \label{eq:qi}
\end{equation}%
where $|i\rangle \equiv |00\dots 1_{i}\dots 00\rangle $ denotes the state
with a spin-wave excitation in the $i^{th}$ micro-ensemble and none in
others. From this expression, we get $\sum_{i}$ $q_{i}/\eta
_{i}=p_{1}\sum_{i}\langle i|\rho _{1}|i\rangle =p_{1}$, so we obtain $p_{1}$
and $p_{0}=1-p_{1}$ from the measured $q_{i}$ and $\eta _{i}$. As an
example, for the case of $3\times 3$ micro-ensemble array, the results of
the measured $q_{i}/\eta _{i}$ are shown in Fig.~3d of the main text.
Meanwhile, we can measure the idler photonic single-excitation population by
not correcting the retrieval efficiency, that is, $p'_1=\sum q_i$, which is just
the sum of the measured probabilities in each of the $N$ modes.

For this simple case, it is also easy to determine the fidelity $F$, which is
measured by the setup shown in Fig.~S3. In Fig.~S3b,
the idler AODs are set to equally and coherently combine the idler modes
from all the $N$ micro-ensembles. If we neglect the small inhomogeneity
in the retrieval efficiencies $\eta _{i}$ and replace $\eta _{i}$ with their
average $\overline{\eta }$, the measured probability $q_{f}$ to record a
photon count from the combined mode in Fig.~S3b in each experimental trial is
just given by $q_{f}=\overline{\eta }\langle W_{N}|\rho _{e}|W_{N}\rangle =%
\overline{\eta }F$, which gives the fidelity as $F=q_{f}/\overline{\eta }$
from the measured quantities $q_{f}$ and $\overline{\eta }$. Later we will
take into account both the contributions of the double excitation probability
$p_{2}$ and the inhomogeneity in $\eta _{i}$ to correct the formula for
the fidelity $F$. The photonic fidelity $F'=q_f$ is just the measured probability
of detector in this fidelity measurement setup.

Now we consider the contribution of the double excitation probability $p_{2}$%
. First we need to measure this small probability $p_{2}$ in our experiment.
The measurement configuration is shown by the supplementary Fig.~S4,
where we split the combined idler photon mode by a 50/50 beam splitter and
detect the three-photon coincidence between the single photon detectors D1,
D2, and D3. We measure the normalized three-photon correlation, defined by
\begin{equation}
\alpha \equiv \frac{q_{1}q_{123}}{q_{12}q_{13}},
\end{equation}%
where $q_{1},q_{12},q_{13},q_{123}$ denote respectively the probabilities of
registering a photon count on detector 1, a coincidence of counts between
detectors 1 and 2, a coincidence between detectors 1 and 3, and a
coincidence between detectors 1, 2 and 3 in each experimental trial. By this
definition, $\alpha $ becomes independent of the detector efficiency and the
transfer efficiency from the spin-wave modes to the photon modes as their
contributions to the numerator and the denominator of $\alpha $ cancel with
each other. The normalized correlation is thus given by%
\begin{equation}
\alpha =\frac{p_{2}\langle 0|a_{W}^{2}\rho _{2}a_{W}^{\dag 2}|0\rangle }{%
p_{1}^{2}|\langle 0|a_{W}\rho _{1}a_{W}^{\dag }|0\rangle |^{2}},
\end{equation}
where $a_{W}$ ($a_{W}^{\dag }$) denotes the annihilation (creation) operator
for the spin-wave modes in the W state.

\begin{figure}[ptb]
\includegraphics[width=3in]{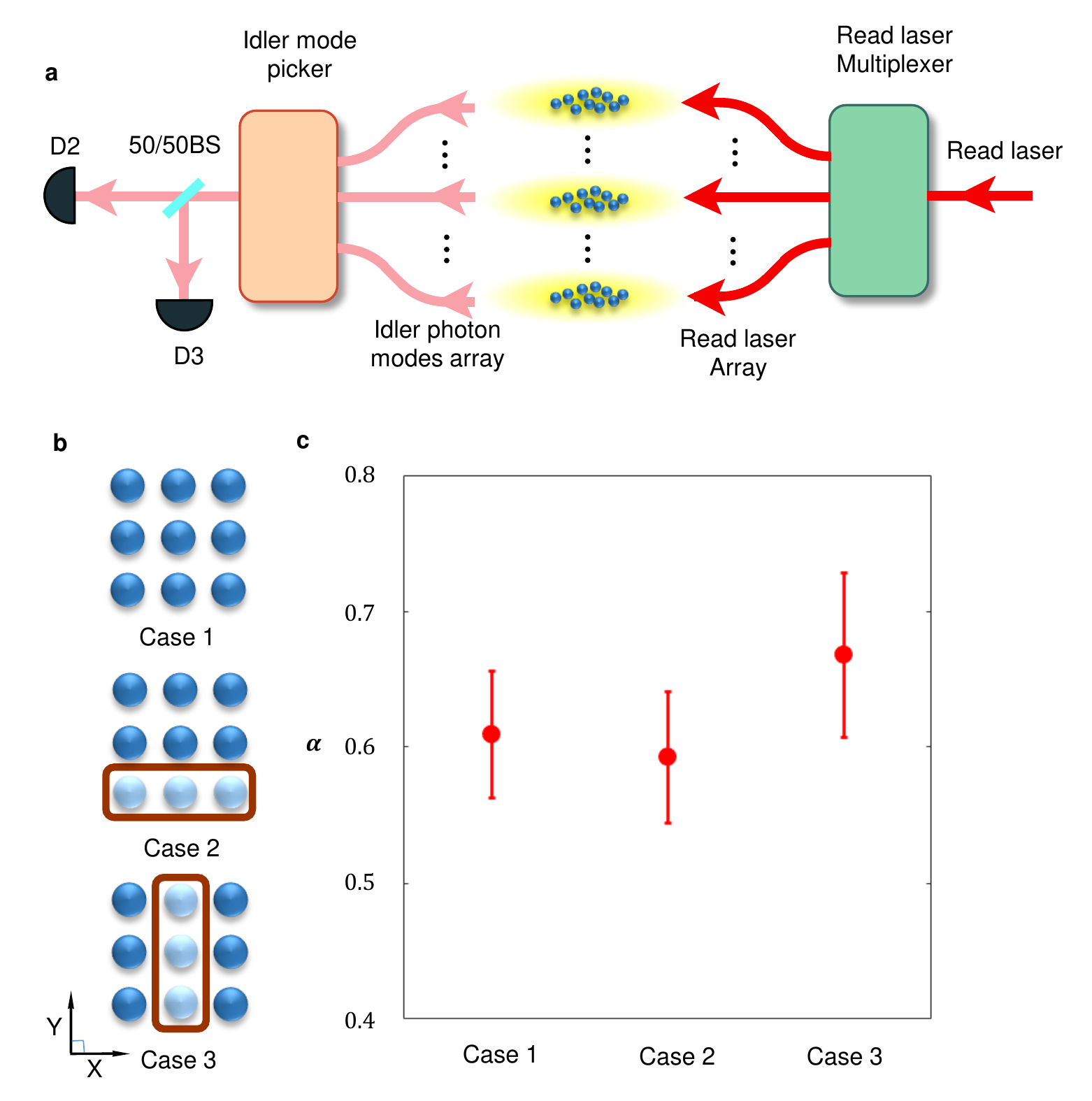}
\caption{ \textbf{Measurement of the three-photon correlation and the double
excitation probability.} \textbf{a,} The detection configuration for
measurement of the double excitation probability $p_2$, where a 50/50 beam
splitter is inserted to split the combined idler mode to detect the double
excitations in this mode. We combine this detection configuration with the
entanglement generation configuration shown in Fig.~S3a,
where the detector D1 registers a signal photon in the combined mode. We
register the three-photon coincidence between the detectors D1, D2, and D3,
and from it construct the normalized three-photon correlation $\protect\alpha
$ and the double excitation probability $p_2$ as explained in the
supplementary text. \textbf{b,c} Test of the three-photon correlation $%
\protect\alpha$ under different measurement bases. We check three
arbitrarily chosen bases here. In case 1, we detect in the full
superposition basis from the $3\times 3$ ensemble array. In case 2 and 3, we
detect in the partial superposition basis from the $2\times 3$ (1st and 2nd
rows) and $3\times 2$ (1st and 3rd columns) ensemble arrays, respectively,
as illustrated in b. The measured three-photon correlations $\protect\alpha$
shown in c are independent of the measurement bases within the experimental
error bar (the error bar corresponds to one standard deviation). Note that
for the test measurement shown in c we have increased the write laser
intensity compared with the one for the W state generation to get a higher
three-photon coincidence rate, so the value of $\protect\alpha$ is also
higher than what we have for the W state preparation experiment. }
\end{figure}

From our excitation configuration for preparation of the W state, the
dominant excitations in $\rho _{1}$ and $\rho _{2}$ should be along the same
spin-wave mode as they come from the driving by the same write beam.
Neglecting small imperfection terms, we can express $\rho _{1}$ and $\rho
_{2}$ approximately as%
\begin{equation}
\rho _{1}\approx a_{e}^{\dag }\rho _{0}a_{e},\text{ }\rho _{2}\approx
a_{e}^{\dag 2}\rho _{0}a_{e}^{2}/2,  \label{eq:12}
\end{equation}%
where $\rho _{0}$ denotes the vacuum for spin-wave excitations and $a_{e}$
denotes the excited spin-wave mode, which in our experiment is quite close
to the mode $a_{W}$ for the W state. The factor of $1/2$ in $\rho _{2}$
comes from normalization $tr(\rho _{2})=1$. Under this approximation, we
have $\langle 0|a_{W}^{2}\rho _{2}a_{W}^{\dag 2}|0\rangle \approx
2\left| [a_{W},a_{e}^{\dag }]\right| ^{4}$, $\langle 0|a_{W}\rho
_{1}a_{W}^{\dag }|0\rangle \approx \left|[a_{W},a_{e}^{\dag }]\right|^2$. So the
normalized correlation reduces to%
\begin{equation}
\alpha \approx \frac{2p_{2}}{p_{1}^{2}}.  \label{eq:13}
\end{equation}%
By measuring the normalized correlation $\alpha $, we thus get a simple
relation between the double-excitation and single-excitation probabilities $%
p_{2}$ and $p_{1}$. (As $\alpha $ is a normalized quantity,
independent of the retrieval efficiency
and the photon loss, for the directly measured idler photon modes,
we still have $\alpha =2p'_{2}/{(p'_{1})^{2}}$.
As $p'_{1}=2q_{12}/q_{1}=2q_{13}/q_{1}$, from the definition of
$\alpha $, we then have $p'_2=2q_{123}/q_{1}$, which
is consistent to what we expect from the definition of
the coincidence counts. In analyzing the experimental data,
we use the formula $p'_2=\alpha ({p'_1})^2/2$ to derive
$p'_2$ from the measured $\alpha$ and $p'_1$, and then
deduce $p'_0$ by $p'_0=1-p'_1-p'_2$.)

Note that with the approximation in Eq.~(\ref{eq:12}), the measured $\alpha $
should be independent of which combinations of the idler photon modes we
detect for the three-photon coincidence if we keep the write beam intensity
fixed (thus $p_{2}/p_{1}^{2}$ fixed). If instead of $a_{W}$, we detect a
different superposition $a_{d}$ of spin-wave modes, the factor of commutator
$[a_{d},a_{e}^{\dag }]$ still cancels in the numerator and the denominator
of $\alpha $. We tested this prediction with the results shown in
Fig. S4b and S4c, where the measured values of correlation $%
\alpha $ for three randomly chosen superposition modes $a_{d}$ are shown.
These values of $\alpha $ remain unchanged within the experimental error bar
although the detected modes $a_{d}$ are quite different. This experimental
test further supports that the approximation in Eq.~(\ref{eq:12}) is valid
for our experiment.

With consideration of the double-excitation probability $p_{2}$, for the
detection of $q_{i}$ with the configuration shown by Fig.~S2,
Eq.~(\ref{eq:qi}) should be replaced by
\begin{equation}
q_{i}=\eta _{i}p_{1}\langle i|\rho _{1}|i\rangle +2\eta _{i}\left( 1-\eta
_{i}\right) p_{2}\langle i,i|\rho _{2}|i,i\rangle +\eta _{i}p_{2}\sum_{j\neq
i}\langle i,j|\rho _{2}|i,j\rangle .
\end{equation}%
where $|i,j\rangle =|00\dots 1_{i}\dots 1_{j}\dots 0\rangle $ denotes the
double-excitation state with spin-wave excitations in the $i$th and $j$th
micro-ensembles. As both $\eta _{i}$ and $p_{2}$ are small for our
experiment, the high-order contribution $\eta _{i}^{2}p_{2}$ is neglibible
in the second term. A summation of the above equation over the index $i$
then gives
\begin{eqnarray}
\sum_{i}q_{i}/\eta _{i} &=&p_{1}+2p_{2}\sum_{i,j,i\leq j}\langle i,j|\rho
_{2}|i,j\rangle   \notag \\
&=&p_{1}+2p_{2}.  \label{eq:15}
\end{eqnarray}%
Combining this equation with $\alpha =2p_{2}/p_{1}^{2}$ and the
normalization $p_{0}+p_{1}+p_{2}=1$, we can determine the population $p_{0}$%
, $p_{1}$ and $p_{2}$ with the measured quantities $\alpha $\ and $%
\sum_{i}q_{i}/\eta _{i}$. The experimental data for $p_{0}$, $p_{1}$ and $%
p_{2}$ in Figs.~3, 4 and 5 of the main text are determined in this way. To
determine the error bar and the confidence intervals, we sample the measured
photon counts and coincidences through the Monte Carlo simulation by
assuming a Poissonian distribution. For each sample of photon
counts/coincidences, we determine the population $p_{0}$, $p_{1}$ and $p_{2}$
through the above equations. The Monte Carlo simulation then gives the
distribution for the parameters $p_{0}$, $p_{1}$ and $p_{2}$, from which it
is straightforward to calculate the error bar and the confidence intervals.
The confidence intervals for all the other quantities in our experiment,
including the fidelity to the W state and the entanglement witness, are
determined in the same way by the Monte Carlo simulation.

Finally, we determine the fidelity $F$ to the W state by taking into account
both of the double excitation probability and the small inhomogeneousity in
the retrieval efficiencies from different micro-ensembles. The experimental
setup to measure $F$ is still given by Figs.~S3. The
idler AODs in Fig.~S3b equally and coherently combine the idler photon modes
from $N$ micro-ensembles. The retrieval efficiency for the $i$th ensemble is
$\eta _{i}$. The total transfer efficiency from a spin-wave excitation in
the $i$th ensemble to a photon click on the idler photon detector is thus
given by $t_{i}=\eta _{i}/N$. Let $T=\sum_{i}t_{i}$ and $t_{i}^{\prime
}=t_{i}/T$. The measurement then corresponds to a projection to the state
\begin{equation}
|W_N^{\prime }\rangle =\sum_{i=1}^{N}\sqrt{t_{i}^{\prime }}|i\rangle ,
\end{equation}%
where we have neglected the unimportant relative phases between the
superposition terms as with appropriate setting of the RF phases in the AODs
the relative phases cancel with each other and they can be absorbed into the
definition of the number state $|i\rangle \equiv |00\dots 1_{i}\dots
00\rangle $. By taking into account the double excitation probability $p_{2}$%
, conditioned on a click on the signal detector (D1), the success
probability to register a photon count on the idler detector in Fig.~S3b is
given by
\begin{equation}
q_{f}=T(p_{1}+2p_{2})\langle W_{N}^{\prime }|\rho _{1}|W_{N}^{\prime
}\rangle ,
\end{equation}%
where the $2p_{2}$ term similarly comes from the contribution of $\rho _{2}$%
, which, due to the small transfer efficiency $t_{i}$, is twice the
contribution of $\rho _{1}$ (same as the derivation made in Eq.~(\ref{eq:15}%
)).

From the measured conditional probability $q_{f}$, we derive a lower bound
on the W state fidelity $F$. As the identity operator $\widehat{I}\geq
|W_{N}^{\prime }\rangle \langle W_{N}^{\prime }|$, we have

\begin{eqnarray}
F &\equiv &\langle W_{N}|p_1\rho _{1}|W_{N}\rangle  \notag \\
&\geq &p_1\langle W_{N}^{\prime }|\rho _{1}|W_{N}^{\prime }\rangle |\langle
W_{N}^{\prime }|W_{N}\rangle |^{2} \\
&=&\frac{p_1q_{f}}{T(p_{1}+2p_{2})}|\langle W_{N}^{\prime }|W_{N}\rangle
|^{2}.  \notag
\end{eqnarray}%
The overlap $|\langle W_{N}^{\prime }|W_{N}\rangle |^{2}=\left(
\sum_{i=1}^{N}\sqrt{t_{i}^{\prime }/N}\right) ^{2}$ and the total transfer
efficiency $T$ are known quantities as all the retrieval efficiencies $\eta
_{i}$ have been calibrated. The single and double excitation probabilities $%
p_{1}$ and $p_{2}$ are determined already from the experimental measurements
described before. From the above equation, we then obtain the lower bound to
the W state fidelity $F$ from the measured $q_{f}$. With this lower bound to
$F$ and the measured values of $p_{1}$ and $p_{2}$, we determine an upper
bound to the entanglement witness $\mathrm{tr}(\mathcal{W}_{k}\rho _{e})$,
which can then be used to verify multipartite entanglement. The measurement
results from the above procedure are summarized in Figs.~3, 4 and 5 of the main
text.

Now we estimate the influence from the higher order excitation to our
results. Because the excitation is a parametric process, we can assume that the excitation number in each
micro-ensemble follows an independent Poisson distribution.
The total excitation number is just their sum, thus still follows a Poisson distribution.
In the experiment, we have measured the ratio of $p_2/p_1$ for the probability to have totally one or two excitations. This ratio is related to the parameter $\lambda$ of the Poisson distribution by
$p_2/p_1=\frac{\lambda^2}{2}/\lambda=\frac{\lambda}{2}$, thus
$\lambda=3.11\%$ $(5.89\%,6.34\%)$ for the $N=9$ $(16, 25)$
case, respectively. The total probability
of higher-order excitations is $p_{i>2}<\frac{p_3}{1-\lambda}\approx p_3$, and
we renormalize $p_0, p_1, p_2, F$ by a factor $\frac{1}{1+p_{i>2}}$. With these
modified $p_0, p_1, p_2, F$, we calculate the corrected witness distribution, and
compare them with the uncorrected values. For all the cases ($N=9,\,16,\,25$) the entanglement depth for the spin-wave state stays the same, with a tiny decline in confidence level of the order $10^{-9}$.

\section{Section S3. Discussion of the experimental noise}

We estimate that the major contribution to the entanglement infidelity in our experiment comes from the double excitation probability $p_2$, which has been analyzed in detail above. For the $p_1$ components in the density operator (the single-excitation components), the factors contributing to the entanglement infidelity include the small imbalance of the multi-path interferometer composed by these micro-ensembles and the residue relative phase between different optical paths caused by the imperfect setting of the compensation RF phases. This imbalance has been taken into account in the above measurement of the entanglement witness. The nonzero $p_0$ (vacuum) component, which has influence on our entanglement witness, is mainly caused by
the imperfections in the heralding process, including imperfect filtering of the write laser pulse and the detector dark counts. The decay of the collective atomic excitation in each micro-ensemble can also contribute to the $p_0$ component. In our previous experiment with the same configuration for the micro-ensembles \cite{pu2017experimental} (except that the separation between the micro-ensembles is slightly larger in this experiment), we have measured that the storage time in each micro-ensemble is about $28\,\mu$s and we expect that the same storage time holds for this experiment. The storage time is mainly limited by the dephasing of the collective mode
caused by the atomic random motion at a temperature of about $30\,\mu$K and the small residual magnetic
field gradient when the magneto-optical trap is shut off. The storage time could be significantly extended if we trap the micro-ensembles with far-off-resonant optical traps and use the clock transition in the $\mathrm{Rb}$ hyperfine levels to store the collective excitation. In our current experiment, after a signal photon is registered, the delay time to retrieve the collective atomic excitation to the idler photon is taken to be $5.8\,\mu$s, which is significantly less than the measured storage time of $28\,\mu$s for each micro-ensemble, so we expect that the decay of the atomic excitation during this delay time only plays a minor role to the vacuum component in this experiment.

In our experiments, there are various sources of noise that contribute to the photon loss. The photon loss has no direct
influence on the entanglement fidelity of our experiment as its effect is factored out in our heralded scheme. Nevertheless, these
sources of noise affect the success probability of our heralded protocol. In the following, we briefly discuss the photon loss
channels (various inefficiencies) in the write and read process of the collective atomic excitation.

In the write process, the signal photon mode is defined by the single mode fiber,
and the coupling efficiency to the fiber is about $80\%$. Together with the single
photon detector efficiency of about $50\%$, the transmission efficiency of
an interference filter about $70\%$ and the efficiencies of two first-order diffractions through
the two signal AODs each about $75\%$, the total success probability of registering a
photon count when a signal photon is emitted is therefore about $16\%$, estimated by a multiplication of the
above efficiencies.

In the read process, the overall retrieval efficiency from an atomic spin-wave excitation
to a photon count registered by the idler photon detector is measured to be about $4\%$, as shown in
Fig.~3c of the main text. This retrieval efficiency includes contributions from the following channels of the photon loss: two successive
first-order diffractions on the idler AOD pair (each of about $75\%$ efficiency), the coupling efficiency of the idler photonic mode into a single mode fiber
(about $80\%$), the transmission of an interference filter (about $90\%$), the coupling efficiency of a fiber beam-splitter (about $80\%$),
the total transmission efficiency of the idler photon in the fiber and in the free space (about $95\%$), the efficiency of the single photon detector
(about $50\%$), and the intrinsic retrieval efficiency from the atomic spin-wave mode to the idler photon mode that couples into the single-mode fiber.
From the above numbers, we estimate that the intrinsic retrieval efficiency is about $26\%$, which is consistent with the estimations in other
atomic ensemble experiments at comparable optical depths \cite{sangouard2011quantum}.

\end{document}